\documentclass[11pt,a4paper]{article}
\usepackage{bm}
\usepackage{graphicx}
\usepackage{graphics}
\usepackage{float}
\usepackage{amsmath}
\usepackage{amssymb}
\usepackage{amsfonts}
\usepackage{latexsym}
\usepackage{epsfig}
\usepackage{dcolumn}
\usepackage{url}

\usepackage[titles]{tocloft} 

\usepackage[francais,english]{babel}
\usepackage{lmodern} 
\usepackage[T1]{fontenc}
\usepackage{pdfsync}
\usepackage[pdftex,bookmarks,colorlinks,breaklinks]{hyperref}
\hypersetup{linkcolor=black,
citecolor=blue,filecolor=red}

\newcommand{\Ecal}{{\mathcal E}}

\newcommand{\half}{{\textstyle \frac{1}{2}}}

\newcommand{\fourth}{{\textstyle \frac{1}{4}}}

\newcommand{\avec}{{\textbf a}}
\newcommand{\bvec}{{\textbf b}}

\newcommand{\pvec}{{\textbf p}}
\newcommand{\qvec}{{\textbf q}}

\newcommand{\Svec}{{\textbf S}}

\newcommand{\Lvec}{{\textbf L}}

\usepackage{epstopdf}

\title{\sf defects in quasicrystals, revisited  \\ I$-$  flips, approximants, phason defects}

\author{\textsf{Maurice Kleman}\footnote{\textsf{kleman@ipgp.fr}} 
 \vspace {15pt}\\
\small Institut de Physique du Globe de Paris $-$ 
 Sorbonne Paris Cité\\
 \small 1, rue Jussieu - Paris cedex 05, France}
 \small \date{March 22, 2012} 
  
\begin{document}
\maketitle

\scriptsize
\tableofcontents
\newpage
\normalsize

\begin{abstract} The recent discovery of {metadislocations} in some periodic complex metallic alloys and of their 'phason' defects has given a new impetus to the study of QC approximant defects. In this paper we emphasize: 1$-$ that approximants differ from a QC by a suitable density of 'flips' only, one per unit cell in the case of Fibonacci approximants; these flips are not topological defects, 2$-$ a flip can split into two 'phason' defects of opposite signs, thus the approximant defects can be studied in a first step as defects of the parent QC. In a companion paper this analysis of QC defects is extended to dislocations; the difference between perfect and imperfect dislocations is emphasized. Imperfect dislocations are the phason defects alluded to above.
\end{abstract}
\section{\sf introduction} \label{int}
${\qquad}$
  My renewed interest in the quasicrystal defect theory starts with the 
discovery by the J\"ulich group of a family of very remarkable line defects present in periodic complex metallic alloys,  \textsf{metadislocations}, which display an extremely small Burgers' vector compared to the unit cell parameter and are attended by specific 'phason' defects, see Ref.~\cite{feuerbacher11b} for a review. These investigations have motivated a reappraisal of the question of approximants and of their defects, see e.g. \cite{engel05,gratias12}, a question previously investigated in many papers, \cite{jaric87,entin88,dmitrienko93,gratias95} to cite a few. It is this author contention that metadislocations enter the framework of quasicrystal defects, without need for new concepts in the defect theory of the solid state, at least in a first step. This paper (I) and the following (II) ref. \cite{kleman13c}, which constitute an extended synthesis of previous published articles \cite{entin88,kleman03a}, are devoted to the definition of approximants as quasicrystals with specific defects (I) and to new considerations on perfect $\bvec_{||}$ and imperfect $\bvec_\bot$ dislocations in quasicrystals (II).\smallskip
 
The sites of a rational approximant are the intersections of a set of atomic surfaces  (i.e. copies of the $d_\bot$-dimensional projection onto E$_{\bot}$ of a cell of the $d$-dimensional hypercubic lattice $\Ecal$) attached to a $d_{||}$-rational cut $\mathrm{E}$ with an irrational $d_{||}$-space $\mathrm E_{||}$ that carries the physical structure; $\mathrm{E}, \ \mathrm{E_{||}}, \ \mathrm{E_\bot} \subset \Ecal$, $d=d_{||}+d_{\bot}$.
 In this way the resulting atomic distances are the same as in the $i$-phase. In \cite{gratias12} 
rational approximants are analytically described as the result of an homogeneous shear of the underlying hyperspace, as in \cite{jaric87}. \bigskip

 
 We have proposed in \cite{entin88} a different way of generating an approximant, which has the advantage of showing directly how the periodic lattice inherits the characteristics of the defects of the parent quasicrystal. The unit cell of an approximant is a piece of the parent QC with a certain number of flips, in fact only one flip in the simplest ones $-$ the Fibonacci approximants $-$  which can conveniently be located at the vertices of the unit cell. We recall that a \textsf{flip} is a local displacement of an atom, which, in the tiling representation of QCs, consists in the shift of a vertex which respects the tile shapes but breaks the tile matching rules; see examples hereunder. This is developed in Sect.~\ref{con}.
 
 In this paper we present a complete view of the nature of rational approximants, how they relate to the parent QCs by the presence of 'flips', and how those flips can split into 'phason' defects, which are nothing else than the 'imperfect' dislocations of the QC, generally loosely attached to the perpendicular component of the Burgers vector $\bvec_\bot$. Flips are not topological defects, so that approximants can be thought of as elastic instabilities of a parent QC; this is indeed the description adopted by Jari\v{c} \& al. \cite{jaric87}. On the other hand flips can split into 'phason' defects of opposite signs, each of them being a true topological defect. Therefore if ever this splitting is favored, and opposite phason defects repulsive, the approximant might be more stable than the parent QC. A detailed study of the 'phason' defects, which are true imperfect dislocations, is made in a companion paper (defects in quasicrystals, revisited II$-$ perfect and imperfect dislocations), in relation with the perfect dislocations $\bvec_{||}$ to which they can be paired. 

\section{\sf construction of a rational approximant} \label{con}
In this paper we employ the classical construction of a QC: the atoms $\{m\}$ are the intersections with an irrational cut E$_{||}\subset\Ecal$ of a set of atomic surfaces.
 We consider in turn the cases: $d_{||}=d_{\bot}=1; \  d_{||}=2,\ d_{\bot}=3; \ d_{||}=d_{\bot}=3$.
\subsection{\sf Fibonacci approximants in one dimension} 
\begin{figure}[t] 
   \centering
   \includegraphics[width=4.5in]{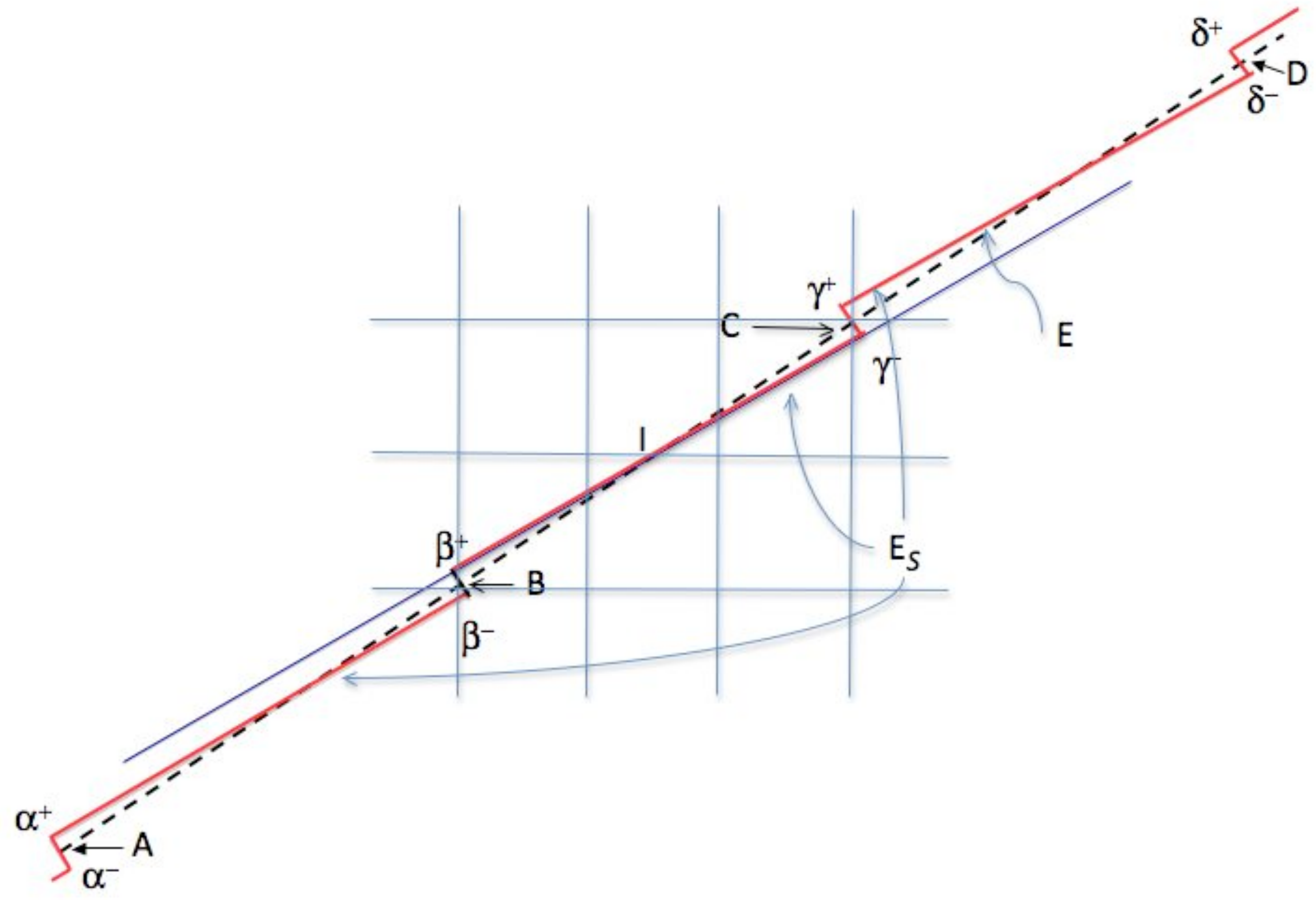} 
   \caption{\scriptsize {Three periods of a Fibonacci approximant $\{f_3, f_4\}$ in the hyperspace $d=2$ (sawtooth cut $\mathrm E_S$). $\dots\alpha^+ \beta^-,\ \beta^+\gamma^-\ \gamma^+ \delta^- \dots$ are irrational segments parallel to $\mathrm E_{||}$, slope $\tau^{-1}$, linked by the 'phason' shifts $\dots\alpha^- \alpha^+, \ \beta^- \beta^+, \ \gamma^- \gamma^+\dots$ perpendicular to E$_{||}$; $\dots$ A, B,C $\dots$ are {\it vertices} of the lattice. $\bf{AB}=\bf{BC}=\dots$ $=\{f_4, f_3\}$. The approximant generated by the sequence of irrational cuts is the same as the approximant generated by E (dashed rational line), slope $f_3/f_4 = 2/3$.}}
   \label{fig1}
\end{figure}

\quad{}
A Fibonacci approximant is represented in a 2D hyperspace by a cut E whose slope  is the ratio of two consecutive Fibonacci numbers $f_n, \ f_{n+1}$. A rational line of slope $p=f_{n}/f_{n+1}$ is the best approximation to the irrational line E$_{||}$ of slope $\tau^{-1}$ ($\tau=(\sqrt 5+1)/2$ is the golden ratio) in the sense that there are no vertices of the square lattice in a double triangle like $\mathrm B \beta^+ (\mathrm I)\gamma^-\mathrm C$, if B and C are vertices of this lattice, see Fig.~\ref{fig1}; the segment BC has slope $p$, the segment $ \beta^+\gamma^-$ has slope $\tau^{-1}$. This result is independent of $n$; besides the area $\sigma$ of the double triangle above is also independent of $n$, $\sigma =  5^{-\half}$, see appendix. 

This has important consequences: a 'phason' shift which moves $\alpha\beta$ to AB does not meet any vertex of the hyperlattice, and thereby there is no {\sf flip} of the atomic sites during this move. This allows for the representation of the cut E of a rational Fibonacci approximant by a sawtooth-like sequence E$_S$ of segments of slope $\tau^{-1}$, linked by 'phason' shifts $\alpha^-\alpha^+,\ \beta^-\beta^+, \cdots$, see Fig.~\ref{fig1}. Thereby as far as we are interested in the construction of a Fibonacci approximant along E, {\it E$_S$ is equivalent to E}.

\begin{figure}[h] 
    \centering
    \includegraphics[width=4.5in]{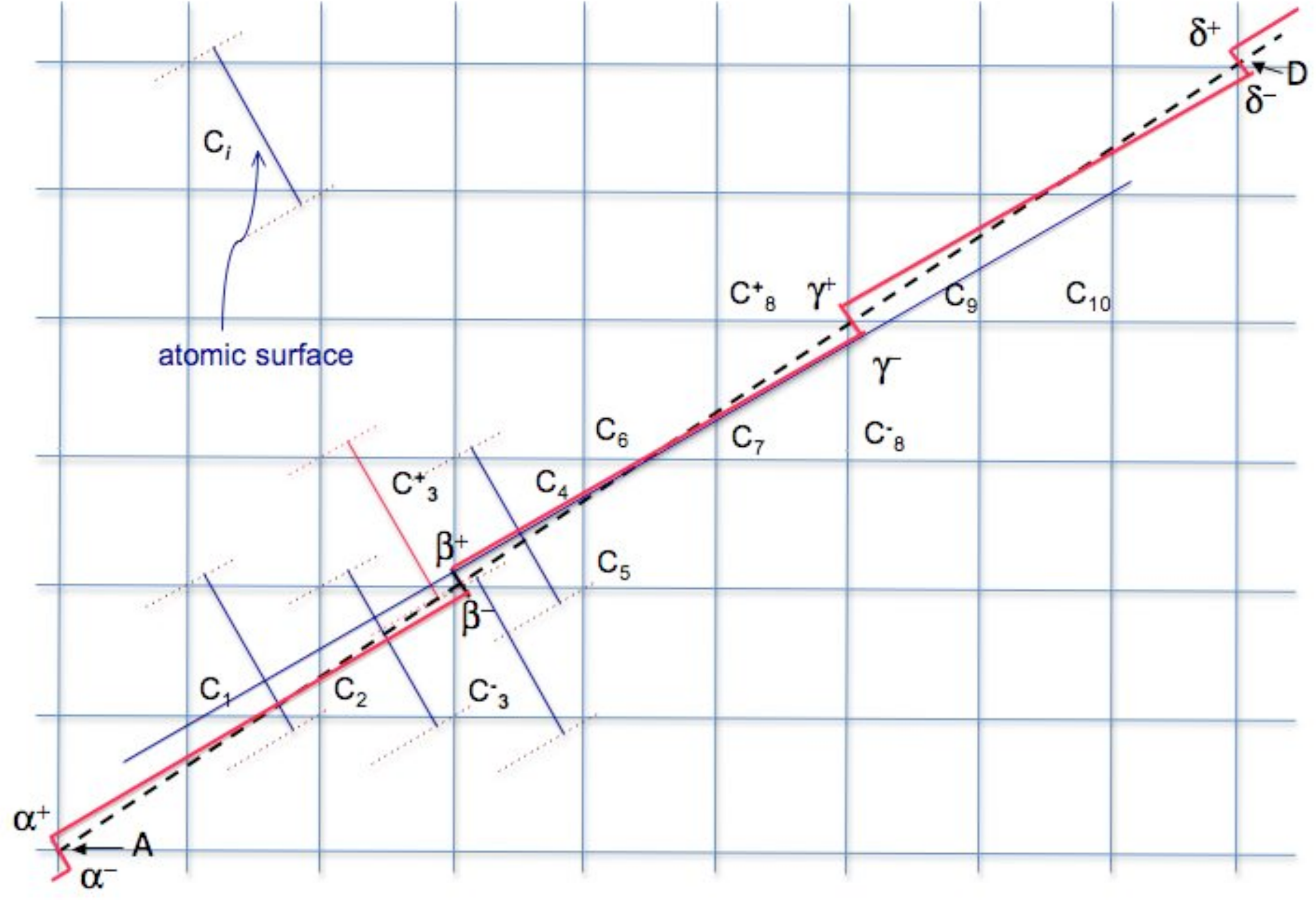} 
    \caption{\scriptsize{The same as Fig.~\ref{fig1}, with some atomic surfaces represented. See text.}}
    \label{fig2}
 \end{figure}

The approximant can be constructed as follows. Attach an atomic surface (a copy of the projection of a cell on the perpendicular space) to each cell center, Fig.~\ref{fig2}, upper left corner; those atomic surfaces that intersect E$_S$ determine a sequence of S (for {\it short}) segments of length $\varsigma$ and L (for {\it long}) segments of length $\ell = \tau \varsigma$, when projected upon a copy of E$_{||}$ . In the case of a Fibonacci approximant, the period in 'real' 1-space consists in $f_{n+2}= f_{n}+ f_{n+1}$ segments, namely $ f_{n}$ Ss and $f_{n+1}$ Ls. 

However this process, when applied to the cut E$_S$ (or E) of Fig.~\ref{fig1}, produces an ambiguity, shown  Fig.~\ref{fig2}. Because of the high symmetry of the E$_S \ +$ E arrangement with respect to the hyperlattice, the atomic surfaces attached to the cells C$_3^-$ and C$_3^+$ both do not intersect E$_S$ and E (or possibly both intersect E$_S$ and E for different atomic surfaces), so that a choice has to be made. It is equivalent to slide E$_S$  to the right (then C$_3^-$ is active) or to the left (then C$_3^+$ is active
)  by a sufficient amount. The sequences of active atomic surfaces are then either 
$\dots$C$_2^{}$C$^-_3$C$_4^{}\dots$ or $\dots$C$^{}_2$C$_3^+$C$^{}_4\dots$, i.e. $\dots$SL$\dots$ or $\dots$LS$\dots$. In fact, whatever the choice that is made, the sequence of S and L is the same, the difference amounting to a shift in the chosen origin of the period; it is $\dots$LSLLS$\dots$ for any sawtooth-cut E$_S$ with $n = 3$, wherever it is in the hyperlattice.

This ambiguity being straightened out, we now show that each period exhibits a flip with respect to a perfect quasi-lattice; the comparison which makes sense is between the sequences carried by E$_{||}$ and those carried by E (or E$_S$). For this purpose let us compare a sequence of active atomic surfaces belonging to E: 
$$\mathrm C_1^{}\mathrm C_2^{}\mathrm C_3^{+}\mathrm C_4^{}\mathrm C_5^{}\mathrm C_6^{}\mathrm C_7^{}\mathrm C_{8}^{+}\mathrm C_{9}^{},$$ which yields a sequence of S \& L segments:
$$\mathrm{LSLLSLSL},$$
and those of E$_{||}$ continued along its segment $\beta\gamma'$:
$$\mathrm C_1^{}\mathrm C_2^{}\mathrm C_3^{+}\mathrm C_4^{}\mathrm C_5^{}\mathrm C_6^{}\mathrm C_7^{}\mathrm C_{8}^{-}\mathrm C_{9}^{},$$ which yields a sequence:
$$\mathrm{LSLLSLLS}.$$

There is therefore a flip associated to this transformation E$_{||}\rightarrow\mathrm E_S$ along the segment BC. More generally the 'phason' shifts $\alpha$A, $\beta$B, $\cdots$, introduce one flip per period, comparing with a local quasilattice carried by the segments AB, BC, $\dots$. One can consider that the two lattice vertices at the two extremities of the triangles, each counting for a $\half$ vertex, amount to the presence of 1 vertex per double triangle. 
Thereby it is appropriate to say that there is one {\sf flip}, and only one, per period of a Fibonacci approximant; notice furthermore that $\sigma$ appears as the natural area that has to be swept by the cut line E$_{||}$ in a 'phasonic' move that produces one flip, and only one.

The term {\sf flip} refers to the fact that two neighboring segments along the Fibonacci sequence are interchanged. In terms of an atomic description, if such neighboring segments are a S and a L segments, the atom separating them has been moved by a distance $\ell-\varsigma = \tau^{-2} \ell$ along the $d_{||}=1$ approximant. 

\subsection{\sf Penrose tiling approximants, $
d=5,\ d_{||}=2,\ d_\bot=3$} 
\quad{}
Entin-Wohlman \& al. \cite{entin88} have shown, using the pentagrid definition of a Penrose tiling, that there is also only one flip in a Fibonacci approximant of such a tiling, this flip being anywhere in the unit cell of the approximant. This unit cell can be chosen as a parallelogram of edges
\begin{equation}
\label{e1}
 \bvec_p = f_{p+1}\Lvec_p+ f_p\Svec_p,\qquad \qquad
\bvec_q =f_{q+1}\Lvec_q+ f_q\Svec_q,
\end{equation} for which $p=q\pm 1$,  or a rhombus for which $p=q$. Here $\Lvec_i$ (resp. $\Svec_i$) are the long (resp. short) diagonals of elementary Penrose tiles, which are either thin (resp. thick) rhombi of angles $\alpha = 36^\circ$ (resp. $\alpha = 72^\circ$).\footnote{We recall that these tiles adjust side by side according to matching rules represented by a fixed arrowing of each type of tile \cite{bruijn81}, see hereunder Fig.~\ref{fig4}.} We have
$$\frac{\bvec_p \cdot \bvec_q}{|\bvec_p| | \bvec_q|} = \cos \alpha,$$
 $\alpha =\angle \{\Lvec_p,\Lvec_q\}$ (or $\alpha=\angle\{\Svec_p,\Svec_q\}$ , $\Svec_p$ is parallel to $\Lvec_p$, $\Svec_q$ to $\Lvec_q$). 

In 5-space, this unit cell belongs to a $d_{||}=2$ rational cut E, with periods: 
\begin{equation}
\label{e2}
 \pvec = \{0,f_{p+1}, f_p,-f_p,-f_{p+1}\},\qquad
\qvec =\{-f_{q+1},0,f_{q+1}, f_q,-f_q\},
\end{equation} for a thin unit cell, and 
\begin{equation}
\label{e3}
 \pvec = \{0,f_{p+1}, f_p,-f_p,-f_{p+1}\},\qquad
\qvec =\{-f_{q},-f_{q+1},0,f_{q+1},f_q\},
 \end{equation} for a thick unit cell.
 
\begin{figure}[h] 
 \centering
    \includegraphics[width=3.5in]{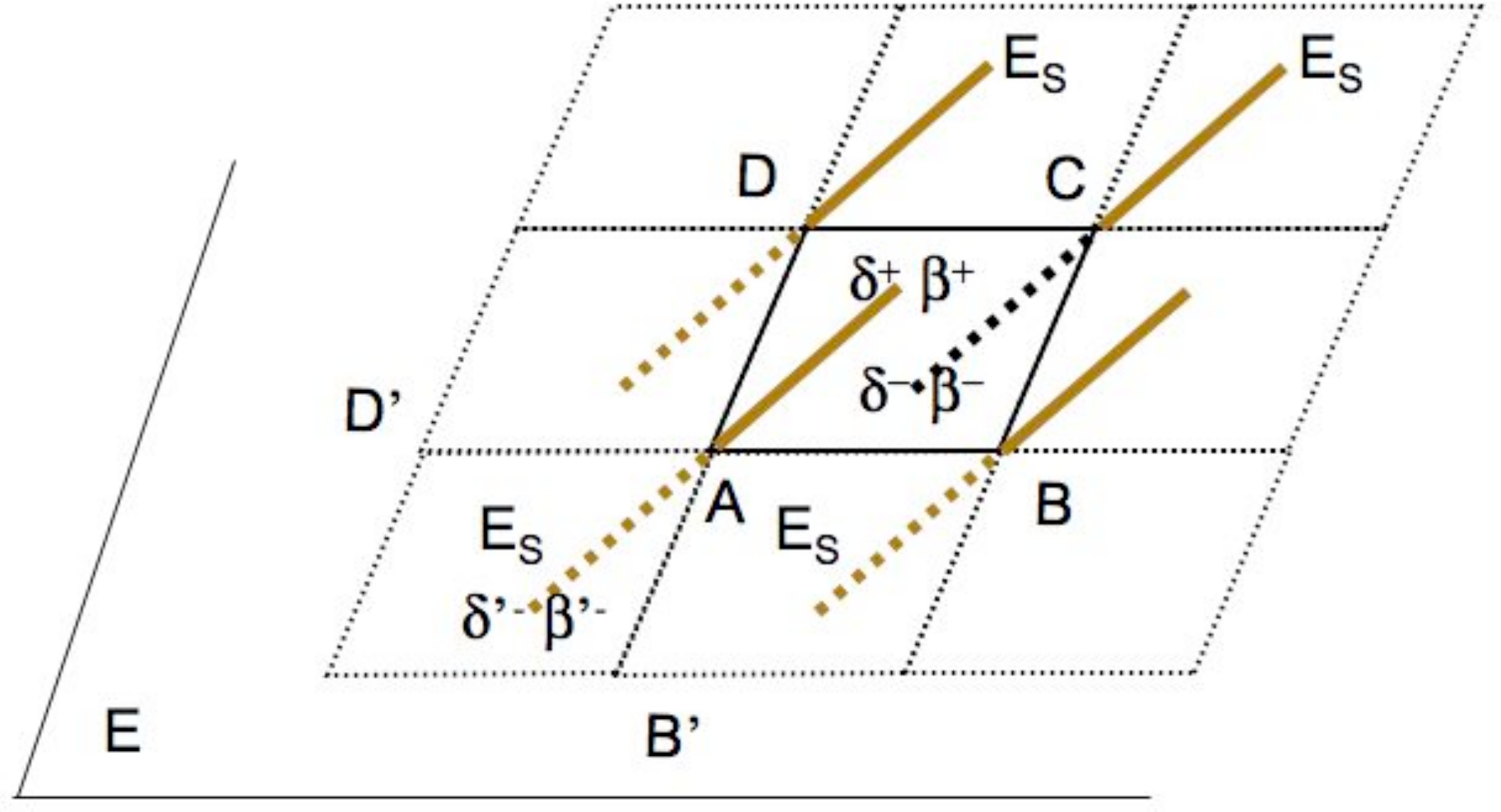} 
 \caption{\scriptsize{Geometric relations between E$_S$, formed of irrational double triangular regions based on E$_{||}$, and the rational cut E. E, which is 2D, and its rhombi-made tiling, is 'correctly' sketched in 2D. Each 2D copy of E$_{||}$ intersects E in only one point (4D geometry), here at vertices of the rhombic tiling, A, B, C, D, $\dots$, and is sketched as a thick segment limited to the projections of $\delta, \ \beta$ neighboring vertices: $\delta^+, \ \beta^+$ are the projections of B, D on the irrational plane attached to A;  $\delta^-, \ \beta^-$ are the projections of B, D on the irrational plane attached to C.}}\label{fig3}
 \end{figure}

 As in the previous case where $d_{||}=1$, one can use instead of this rational cut an
irrational sawtooth-cut E$_S$ which is sketched Fig.~\ref{fig3} for the case $p = q$.
   The rational cut E and the irrational cut E$_{||}$ both belong to the 4 dimensional subspace $\Ecal_4 \in \Ecal_5$, orthogonal to the five-fold axis $\{1,1,1,1,1\}$, of equation $\sum_\imath x_\imath =\gamma = 0, \ \imath = 1,\dots,5$; thus all the operations belong to the same class of {\sf local isomorphism} (LI) in the sense of \cite{socolar86b}. We recall that other LI classes (other values of $\gamma$) correspond to Penrose tiling with different (and more complex) arrowing rules \cite{pavlovitch87}. The generic case $\gamma\neq 0$ will not be considered here.
      
    The 4D sawtooth-cut is constructed, by generalizing the 2D case, as follows. In $\Ecal_4$, E and any copy of E$_{||}$ intersect, generically, only in one point. We choose the vertices of the approximant in the E plane to be such intersections. Consider the vertex A in Fig.~\ref{fig3}; E$_{||}$ in A provides an element of E$_S$ limited by a double triangle ${\delta'}^{-}{\beta'}^{-}$(A)$\delta^+\beta^+$: in A$\delta^+\beta^+$, $\delta^+$ and $\beta^+$ are the projections of B and D on the copy of E$_{||}$ in A $-$ thus \textbf{B}$\bm\beta^+$ and \textbf{D}$\bm\delta^+$ are 'phason' displacements $-$; in A${\delta'}^{-}{\beta'}^{-}$, ${\delta'}^{-}$ and ${\beta'}^{-}$ are the projections of B' and D' on the same copy of E$_{||}$ $-$ \textbf{B}$'\bm{\beta'}^{-}$ and \textbf{D}$'\bm{\delta'}^{-}$ being also 'phason' displacements.  The 4D elements of the sawtooth-cut meeting in A are the simplices ABD$\delta^+\beta^+$ and AB'D'${\delta'}^{-}{\beta'}^{-}$ with a common vertex A, which generalize the 2D elements in Fig.~\ref{fig2}, the double simplex ( a double triangle in 2D) $\mathrm B \beta^+ (\mathrm I)\gamma^-\mathrm C$, with I in common.

As above, there is no vertex of the hypercubic lattice inside the 4D volumes of the simplices ABD$\delta^+\beta^+$ and AB'D'${\delta'}^{-}{\beta'}^{-}$. 
 Each vertex being common to four rhombi in the E plane, there is in fact 4$\times\fourth = 1$ vertex in the ABCD rhombus, and by the same extension as in the 1D case, 1 vertex per rhombus, thus one flip per unit cell in the Fibonacci approximant of a Penrose tiling.
 
 The flips of a Penrose tiling are well-known objects. We revisit them in the next section, in relation with the notion of stacking fault, i.e. a Penrose rhombus edge that does not obey the matching rules.
 
 \subsection{\sf 3D approximants}
 \qquad
 The same arguments can be developed for {3D approximants of $i$-phases}, employing 6D simplices that form a sawtooth. We do not have constructed such geometries, but there is no doubt they exist, and that they yield Fibonacci approximants with one flip per unit cell. A phason singularity is now a face of a Mackay rhombohedron that does not obey the matching rules, in relation with the presence of flips.

\section{\sf about phason defects}\label{pd}
${\qquad}$
 \begin{figure}[h] 
    \centering
    \includegraphics[width=3.3in]{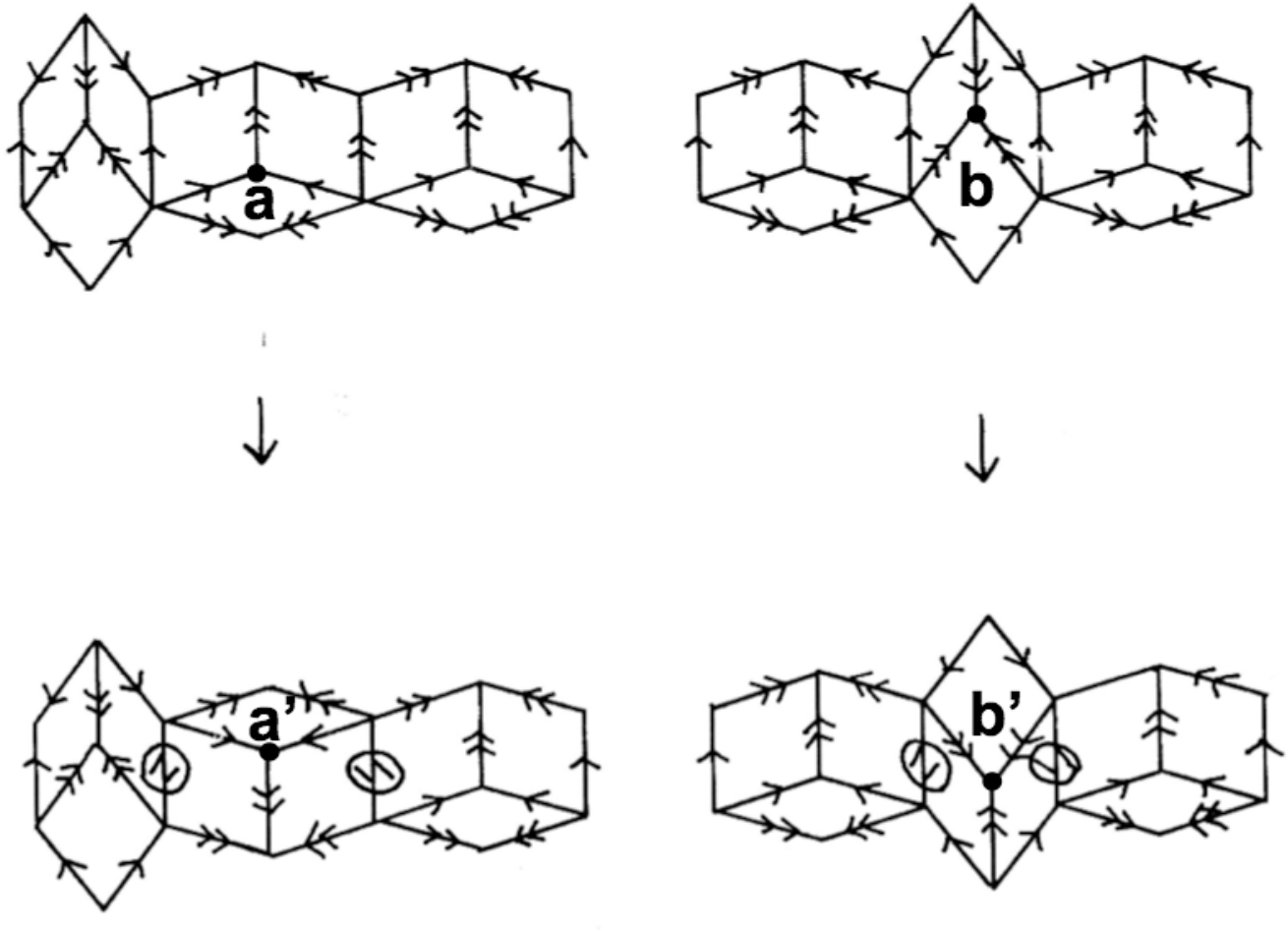} 
    \caption{\scriptsize{Flips $\avec\rightarrow\avec'$ and $\bvec\rightarrow\bvec'$ accompanied by two 'phason' singularities (or mismatches) of opposite signs. The two singularities can diffuse apart along the row of hexagons by a sequence of local flips. Adapted from Fig. 3 in \cite{kleman03a}.}}
    \label{fig4}
 \end{figure}
 A \textsf{flip} is the sum of two singularities, each of them akin to a stacking fault bordered by  an imperfect dislocation dipole. Let us summarize what this statement implies:

$-$ In fig.~\ref{fig4} are sketched two flips in a Penrose tiling that bring the atoms in positions $\bold a,\ \bold b$ to the positions 
$\bold a',\ \bold b'$. Because a flip can be
erased by a movement of the atoms bringing them back to their previous positions, the two singularities (the marked edges on the figure) that accompany it can be thought of as being of opposite signs; it is equivalent to refer to a flip or to these two singularities. The marked edges do not satisfy the matching rules $-$ here represented by the de Bruijn's arrowing \cite{bruijn81} $-$  in a Penrose tiling of isomorphism class $\gamma = 0$. These {\sf mismatches} are the singularities we are alluding to,

$-$ the two mismatches carried by the flip can diffuse apart \cite{kalugin93} by a succession of flips, independently one from the other. Therefore it makes sense to consider a mismatch as a {\sf'phason' singularity} {\it per se}, since it cannot be eliminated by a local movement of the atoms,
 
 $-$ mismatches are true {\sf stacking faults} in the sense of the condensed matter defect theory of defects. The Penrose tiling case is illustrated fig.~\ref{fig5}. The shift that characterizes the fault  is a displacement $\bvec^*$ equal to the flip, fig.~\ref{fig5}c). Figures \ref{fig5} a), b) make explicit the construction of the dislocation dipole $\mathrm L_1, \ \mathrm L_2$, from which one deduces its Burgers vector $\bvec$. The shift $\bvec^*$ is a part of $\bvec$ by virtue of this construction; it is this shift that constitutes the {\sf imperfect} feature of the dislocation. This will be more thoroughly discussed in (II). In 3D, matching faults are made of a 2D fault ({\sf an \emph{imperfect} dislocation}),
 \begin{figure}[t]
\centering
\includegraphics[width=4.5in]{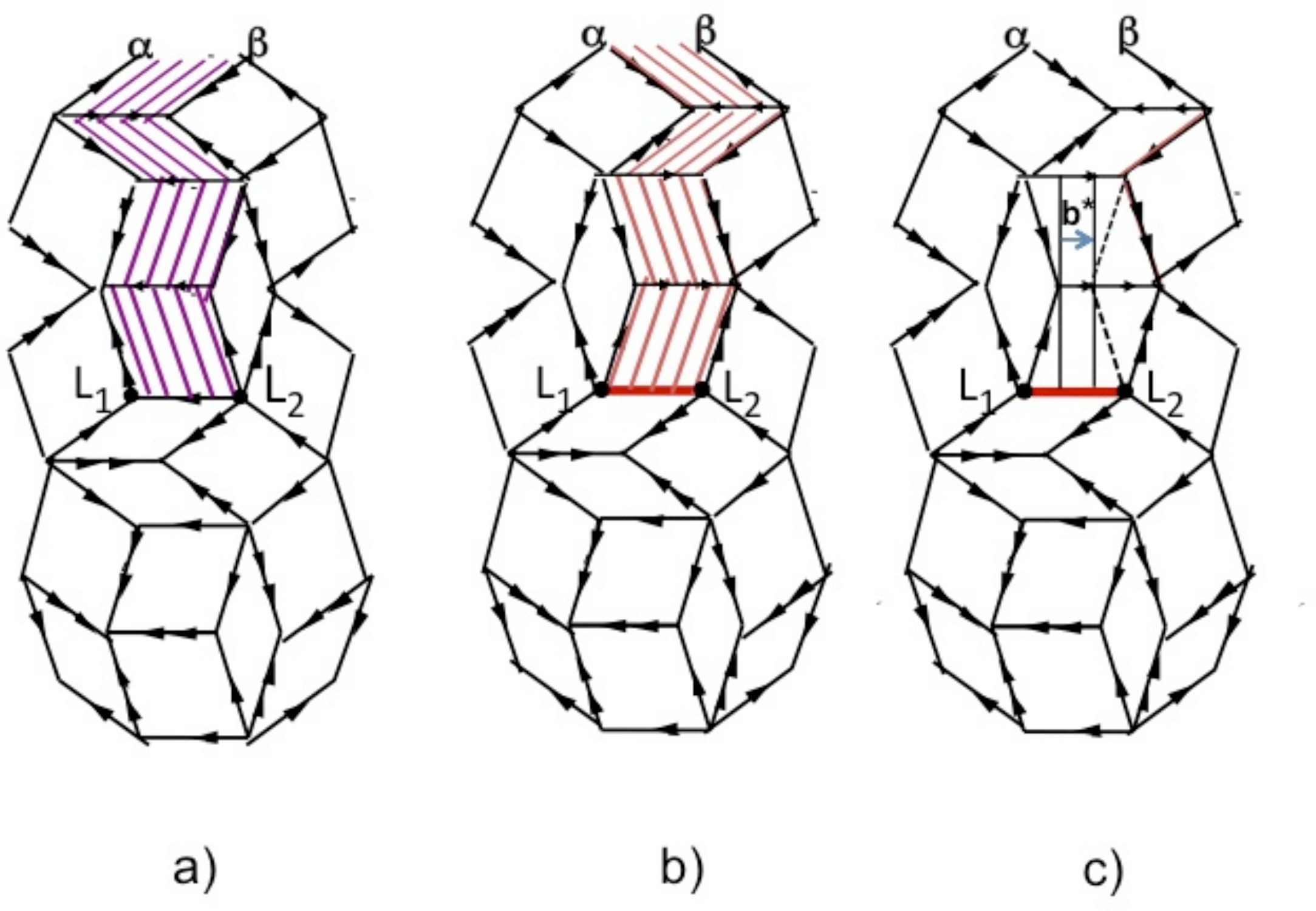}
\caption{\scriptsize{Matching fault in a Penrose tiling $\gamma=0$. A phason singularity is akin to an imperfect dislocation dipole: a) construction of a dislocation of Burgers vector L$_1$L$_2$ ($\bvec= \{1,0,0,0,0\}$ in the hyperspace), with a zigzaging cut surface along L$_1\alpha$; after matter removal (blue) the two zigzag lines L$_1\alpha$, L$_2\beta$ are glued. The plastic deformation thus introduced (not drawn) is relaxed by a dislocation of opposite Burgers vector in L$_2$, implying matter addition (red). b) Final result: L$_1$L$_2$ is a stacking fault (arrowing undefined) bordered by an imperfect dislocation dipole:  $\bvec$ does not belong to the set of $\bvec$ Burgers'vectors. c) The dotted lines show the perfect configuration in contact with the matching fault, before any flip action; $\bvec*$ is the shift of the vertex that defines the matching fault; it is a part of $\bvec$. Adapted in part from Fig. 4 in \cite{kleman03a}.} }
\label{fig5}
\end{figure}


$-$ stacking faults (also called matching faults) are therefore singularities {\it per se}, and as such can be classified by the methods of the topological theory of defects \cite{kleman90,misir95},

\section{\sf discussion}  
\quad
There is some confusion in the literature about the concept of {\it phasons}. We have to distinguish between {\it flips} and {\it topological defects}. A flip is not a topological defect, since an antiflip (exerted on the same atom or cluster) restores the perfect structure. The topological defects we are alluding to are stacking faults, with their usual meaning; because they have been thought of in the wake of the notion of mismatches $-$ a concept that is typical of QCs $-$ they can also be called matching faults. I have used both terms, without making a distinction. Two opposite topological defects can collapse to form a flip.\smallskip

The fact that one finds experimentally the same $\bvec_{||}$s in the approximants and the parent QC \cite{feuerbacher11b} justifies the representation of an approximant in the hyperspace by an saw-tooth-like {\it irrational} cut rather than by a continuous rational cut.

Also, since $\bvec_{||}$ measures a distance between atomic species in a Mackay cluster, one can infer that the cluster shape and scale in the QC and in the approximant are the same. This was anticipated in Entin-Wohlman \& al. \cite{entin88}.
The question therefore arises whether this can be given a meaning in terms of electronic stability rules, applied respectively to the QC and to the approximant \cite{friedel88}? In a sense an approximant might be nothing else than a modification of a QC at practically constant energy. Notice that one does not know of any faceted crystal of an approximant, whereas quasicrystals often grow faceted.

\section*{\sf acknowledgments} I thank V. Dmitrienko, J. Friedel, G. Martin and J.-P. Poirier for useful discussions.

This is IPGP contibution \# 3383

\section*{\sf appendix A}
  \noindent {\small {\sf Calculation of the length $\beta^- \beta^+$}. \normalsize

 We use the Binet's formula, namely
$$f_n \sqrt 5= [\tau^n -(-1)^n\tau^{-n}].$$
The angle $w= \angle{\beta^- \gamma^- \beta^+ } = v-u$, where $u$ and $v$ are the angles of the directions $\mathrm E$ and $\mathrm E_{||}$ with the abscissa, namely $\tan u= f_n/f_{n+1}, \ \tan v = \tau^{-1}$. Thus:

\noindent $\tan w = ({\tan v-\tan u})/({1+\tan u\tan v})$ reads, using the formula above:
$$\tan w = (-1)^{n+1}\tau^{-2n-1}.$$
From that expression one gets 
$\sin^2 w={\tau^{-2n-1}}/ ({f_{2n+1}}\sqrt 5).$ 
Since
$(\beta^- \gamma^-)^2 = (f_n^2 +f_{n+1}^2)\sin^2 w\equiv f_{2n+1}\sin^2 w,$ one eventually gets:
$$\beta^- \beta^+  =(\sqrt 5 \ \tau)^{-\half} \tau^{-n}. $$

\noindent{\sf Calculation of the area of the triangle $\beta^- \gamma^- \beta^+ $}. 

This is a right triangle, and the double of its area is $\sigma = (\beta^-  \beta^+ )\times (\beta^+ \gamma^-)$, hence:
$$\sigma =  f_{2n+1}|\sin w \cos w| \equiv  f_{2n+1}\frac{|\tan w|}{1+\tan^2w} = \frac{1}{\sqrt 5}\approx 0.447214,$$ which is independent of $n$. The Binet's formula is useful in the course of the demonstration of this result.

\newpage
\scriptsize

\bibliographystyle{unsrt}


\begin{thebibliography}{10}

\bibitem{feuerbacher11b}
M.~Feuerbacher and M.~Heggen.
\newblock {Metadislocations}.
\newblock {\em Dislocations in Solids}, 16:110--170, 2011.
\newblock edited by J. P. Hirth and L. Kubin.

\bibitem{engel05}
M.~Engel and H.-R. Trebin.
\newblock {A uniform projection formalism for the Al-Pd-Mn quasicrystals
  $\Xi$-approximants and their metadislocations.}
\newblock {\em Phil. Mag.}, 85:2227--2247, 2005.

\bibitem{gratias12}
D.~Gratias, M.~Quiquandon, and D.~Caillard.
\newblock Geometry of metadislocations in approximants of quasicrystals.
\newblock {\em Phil. Mag. Lett.}, 2012.
\newblock DOI:10.1080/14786435.2012.706372.

\bibitem{jaric87}
M.~V. Jari\'c and U.~Mohanty.
\newblock {"{M}artensitic" instability of an icoshedral quasicrystal}.
\newblock {\em Phys. Rev. Lett.}, 58:230--233, 1987.

\bibitem{entin88}
O.~Entin-Wohlman, M.~Kleman, and A.~Pavlovitch.
\newblock {Penrose tiling approximants}.
\newblock {\em J. Phys. France}, 48:587--598, 1988.

\bibitem{dmitrienko93}
V.~E. Dmitrienko.
\newblock New approaches to the construction of quasicrystals and their cubic
  approximants.
\newblock {\em J. Non-Crystall. Solids}, 153 \& 154:150--154, 1993.

\bibitem{gratias95}
D.~Gratias, A.~Katz, and M.~Quiquandon.
\newblock {Geometry of approximant structures in quasicrystals}.
\newblock {\em J. Phys.: Condens. Matter}, 7:9101--9125, 1995.

\bibitem{kleman13c}
M.~Kleman.
\newblock Defects in quasicrystals, revisited: {II}$-$ perfect and imperfect
  dislocations.
\newblock 2013.
\newblock submitted.

\bibitem{kleman03a}
M.~Kleman.
\newblock Phasons and the plastic deformation of quasicrystals.
\newblock {\em Eur. Phys. J. B}, 31:315--325, 2003.

\bibitem{bruijn81}
N.G. de~Bruijn.
\newblock Algebraic theory of {P}enrose's non-periodic tilings of the plane.
  {II}.
\newblock {\em Kon. Nederl. Akad. Wetensch. Proc. Ser. A (= Indag. Math.)},
  84:53--66, 1981.

\bibitem{socolar86b}
J.~E.~S. Socolar and P.~J. Steinhardt.
\newblock {Quasicrystals. II. {U}nit-cell configurations}.
\newblock {\em Phys. Rev. B}, 34:617--647, 1986.

\bibitem{pavlovitch87}
A.~Pavlovitch and M.~Kleman.
\newblock Generalized {P}enrose tilings: structural properties.
\newblock {\em J. of Phys. A}, 20:687--702, 1987.

\bibitem{kalugin93}
P.~A. Kalugin and A.~Katz.
\newblock {A Mechanism for Self-Diffusion in Quasi-Crystals}.
\newblock {\em Europhys. Lett.}, 21:921--926, 1993.

\bibitem{kleman90}
M.~Kleman.
\newblock Topology of the phase in aperiodic crystals.
\newblock {\em J. Phys. (Paris)}, 51:2431--2447, 1990.

\bibitem{misir95}
T.~Sh. Misirpashaev.
\newblock {Phase Defects and Order Parameter Space for Penrose Tilings}.
\newblock {\em J. Phys. I France}, 5:399--407, 1995.

\bibitem{friedel88}
J.~Friedel.
\newblock {Do metallic quasicrystals and associated {Frank and Kasper} phases
  follow the {Hume-Rothery} rules?}
\newblock {\em Helvetica Physica Acta}, 61:538--556, 1988.

\end{thebibliography}
\end{document}